\documentclass[final,3p,times,sort&compress]{elsarticle}
\usepackage{amsmath,amssymb,ulem,graphicx,booktabs,rotating,marginnote}
\usepackage{citesort}
\graphicspath{{pics/}}
\usepackage{hyperref}
\hypersetup{
colorlinks=true,urlcolor=blue,
	linkcolor = blue,
        urlcolor  = blue,
        citecolor = blue,
        anchorcolor = blue,
}
\hypersetup{pdftitle={Chiral Extrapolation of the Sigma Resonance},pdfsubject={Manuscript}} 

\begin{document}
\begin{frontmatter}
\title{Chiral Extrapolation of the Sigma Resonance}

\author[GWU,JLAB]{Michael~D\"oring}
\author[GWU]{Bin~Hu}\ead{binhu@gwmail.gwu.edu}
\author[GWU]{Maxim~Mai}\ead{maximmai@gwu.edu}

\address[GWU]{Department of Physics, The George Washington University, 725 21$^{\rm st}$ St. NW, Washington, DC 20052, USA}
\address[JLAB]{Thomas Jefferson National Accelerator Facility, 12000 Jefferson Avenue, Newport News, VA 23606, USA}

\begin{abstract} 
We analyze recent results on isoscalar $\pi\pi$ scattering from a $N_f=2+1$ lattice simulation by the HadronSpectrum collaboration by re-summing the two-flavor chiral scattering amplitude of the next-to-leading order in the so-called inverse amplitude method. The lattice data can be well extrapolated to the physical pion mass. We also find that both $I=0$ and $I=1$ lattice data can be described simultaneously for pion masses up to $M_{\pi}=236$~MeV.
\end{abstract}

\end{frontmatter}


\section{Introduction}\label{sec:intro}

The isoscalar $\pi\pi$ scattering amplitude has attracted much interest due to the existence of the broad $f_0(500)$ resonance relatively close to threshold. The resonance, called $\sigma$ in the following, is difficult to distinguish from a structureless background in the $\delta_{00}$ partial wave and cannot be described by a Breit-Wigner resonance. Additionally, there exists a zero in the amplitude below threshold which contributes to the uncommon lineshape of the resonance~\cite{Adler:1964um, Adler:1965ga}. Due to these obstacles a reliable determination of the pole position of the $\sigma$ was difficult until the use of Roy equations together with Chiral Perturbation Theory resulted in a very precise extraction of the pole far in the complex plane~\cite{Caprini:2005zr,GarciaMartin:2011jx}.

Calculating the isoscalar partial wave amplitude from first principles of QCD has been a challenge. Many lattice QCD simulations have been performed in recent years measuring phase shifts in the isovector channel of $\pi\pi$ scattering~\cite{Aoki:2007rd, Lang:2011mn, Bali:2015gji, Guo:2016zos, Aoki:2011yj, Bulava:2016mks, Dudek:2012xn,Wilson:2015dqa}. However, in the isoscalar channel, despite pioneering simulations in the past~\cite{Alford:2000mm, Prelovsek:2010kg, Fu:2013ffa}, phase shifts were never determined due to disconnected quark diagrams among other problems. Only recently the challenging task of measuring the isoscalar channel has been performed by the HadronSpectrum collaboration~\cite{Briceno:2016mjc} at $M_\pi=236$~MeV and $M_\pi=391$~MeV. The phase-shifts in this sector were extracted using the L\"uscher framework~\cite{Luscher:1990ux} in combination with moving frames~\cite{Rummukainen:1995vs}. 

The purpose of the present paper is to extrapolate the phase shifts from Ref.~\cite{Briceno:2016mjc} to the physical point using Chiral Perturbation Theory (ChPT). As a low-energy effective theory of QCD~\cite{Gasser:1984gg} the latter reconciles correlation functions measured in setups with different quark masses. However, in the strict perturbative sense of such an effective theory the radius of convergence is limited, see e.g. Refs.~\cite{Leutwyler:2015jga, Durr:2014oba}. Further, unitarity is fulfilled only up to a given order in such an expansion. On the other hand, imposing unitarity can constrain the pole positions of the $\sigma~(I=L=0)$ and $\rho~(I=L=1)$ resonances from the low-energy $\pi\pi$ scattering amplitude as shown in Ref.~\cite{Colangelo:2001df}, see also the recent exhaustive review~\cite{Pelaez:2015qba} on the properties of the $\sigma$ resonance. This is the motivation behind numerous methods, developed over the last two decades, to ensure elastic unitarity when starting from a given ChPT amplitude at a given order, see, e.g., Refs.~\cite{Oller:1997ti, Oller:1998hw, GomezNicola:2007qj, Hanhart:2008mx, Nebreda:2010wv, Nebreda:2011di}. Quite recently, the quark-mass dependence of the sigma pole was studied in a different approach~\cite{Bruns:2016zer}, employing a Resonance Chiral Lagrangian for the $\sigma$ resonance. 
We wish to emphasize that the lattice data on isoscalar phase shifts were extracted directly from the plots of the preprint~\cite{Briceno:2016mjc}, i.e. we rely on data that have not yet gone through peer-review at the date of submission of this manuscript. Furthermore, we take the correlations among the energy eigenvalues in the $\rho$ channel into account, whereas those of the $\sigma$ channel are not included in the present analysis. 

In Sec.~\ref{sec:form} we recall the basic features of the modified inverse amplitude method. Then, in Sec.~\ref{sec:pred} we compare the extrapolation of this and two other methods of unitarized chiral perturbation theory (UChPT) to the lattice QCD phase shifts of Ref.~\cite{Briceno:2016mjc}. In Sec.~\ref{sec:fit} the phase shifts of Ref.~\cite{Briceno:2016mjc} themselves are analyzed and extrapolated back to the physical point using 1-loop UChPT. As the UChPT approach should be in principle capable to describe various quantum numbers in $\pi\pi$ scattering simultaneously, we also perform fits to the isoscalar and isovector channels at the same time, using the isovector lattice QCD phase shifts by the HadronSpectrum collaboration~\cite{Dudek:2012xn,Wilson:2015dqa} in addition to their isoscalar ones. Of special interest are the values of the low-energy constants and their consistency with standard ChPT values that is discussed in depth in the same section. Finally, we discuss the properties of the $\sigma$ resonance and in particular its coupling to two pions as a function of the pion mass.

\section{Formalism}\label{sec:form}

To establish a reliable connection between scattering amplitudes at different pion masses, but also to address the non-perturbative regime of $\pi\pi$ scattering, we rely in the following on the so-called inverse amplitude method (IAM)~\cite{Truong:1988zp}. It has been  shown to be very successful in describing all experimental data on $\pi\pi$ scattering~\cite{Hanhart:2008mx,GomezNicola:2007qj}. In the following, we briefly discuss the main properties of the IAM referring for further details and derivation techniques to the original publications, i.e., Refs.~\cite{Hanhart:2008mx,Pelaez:2006nj,Pelaez:2010fj,Truong:1988zp,GomezNicola:2007qj,Nebreda:2010wv,Nebreda:2011di}.

The inverse amplitude method is based on the leading (LO) and next-to-leading order (NLO) chiral amplitudes projected to a specific isospin ($I$) and partial wave ($L$), namely $T_2^{IL}(E)$ and $T_4^{IL}(E)$, respectively. A unitary scattering amplitude $T_{IAM}^{IL}(E)$ is derived using dispersion relations,
\begin{align}\label{eq:IAM}
 T_{IAM}^{IL}(E)=\frac{(T_2^{IL}(E))^2}{T_2^{IL}(E)-T_4^{IL}(E)}\,,
\end{align}
which indeed reproduces the usual chiral expansion up to the next-to-leading order. At the leading order, the chiral amplitude is a function of energy, Goldstone-boson mass, $M^2=B(m_u+m_d)$, and pion decay constant in the chiral limit, $F_0$, only. The amplitude $T_4^{IL}$, however, involves two \footnote{More constants are involved when three-flavor chiral perturbation theory is considered.} low-energy constants (LECs) $\bar l_1$ and $\bar l_2$. Two additional low-energy constants $\bar l_3$, $\bar l_4$ enter the NLO chiral amplitude when replacing the above mass and decay constants by their physical values using one-loop results from Ref.~\cite{Gasser:1983yg},
\begin{align}\label{eq:fpi}
M_\pi^2=M^2\left(1-\frac{M^2}{32\pi^2F_0^2}\bar l_3\right)\text{~~~and~~~}
F_\pi=F_0\left(1+\frac{M^2}{16\pi^2F_0^2}\bar l_4\right)\,.
\end{align}
Note that the $\bar l_i$ are scale-independent constants, depending only on the parameters of the underlying theory - the quark masses. Therefore, they are of no use for extrapolation of the scattering amplitude to different quark masses. However, they are related to the scale dependent, quark-mass independent renormalized LECs via
\begin{align}
l_i^r=\frac{\gamma_i}{32\pi^2}\left(\bar l_i+\log \frac{M^2}{\mu^2}\right) \quad\text{with }\quad \gamma_1=\frac{1}{3}\,,\gamma_2=\frac{2}{3}\,,\gamma_3=-\frac{1}{2}\,,\gamma_4=2\,.
\end{align}
Hence, for a fixed scale $\mu$ one can determine the renormalized LECs and then make predictions for the two-particle scattering for a setup with a different physical pion mass. In the course of this work we will fix the scale to $\mu=770$~MeV.

It is further important to note that chiral symmetry dictates that the isoscalar $\pi\pi$ amplitude vanishes at some energy below threshold. Therefore, the IAM scattering amplitude~\eqref{eq:IAM} becomes singular in this energy region. A modification of IAM (mIAM) was derived in Ref.~\cite{Hanhart:2008mx} to overcome this so-called Adler Zero singularity. Using dispersion relations, it was argued that the modification amounts in adding the following term according to
\begin{align}\label{eq:mIAM}
T_{IAM}&=\frac{(T_2)^2}{T_2-T_4+A_m(s)}\,,\nonumber\\
A_m(s)&=T_4(s_2)-\frac{(s_2-s_A)(s-s_2)\left(T_2'(s_2)-T_4'(s_2)\right)}{s-s_A}\,,
\end{align}
where the indices $I$ and $L$ are suppressed for brevity, and $s_A$ and $s_2$ are the zeros of $T_2(s)-T_4(s)$ and $T_2(s)$, respectively. As a matter of fact, this modification only alters the isoscalar amplitude and in the isovector channel the chiral amplitude vanishes at threshold at every order. Thus, the modification function $A_m(s)$ in Eq.~\eqref{eq:mIAM} is exactly zero in this channel.

\section{Predictions of lattice results}\label{sec:pred}

\begin{figure}[t]
\includegraphics[height=5.8cm]{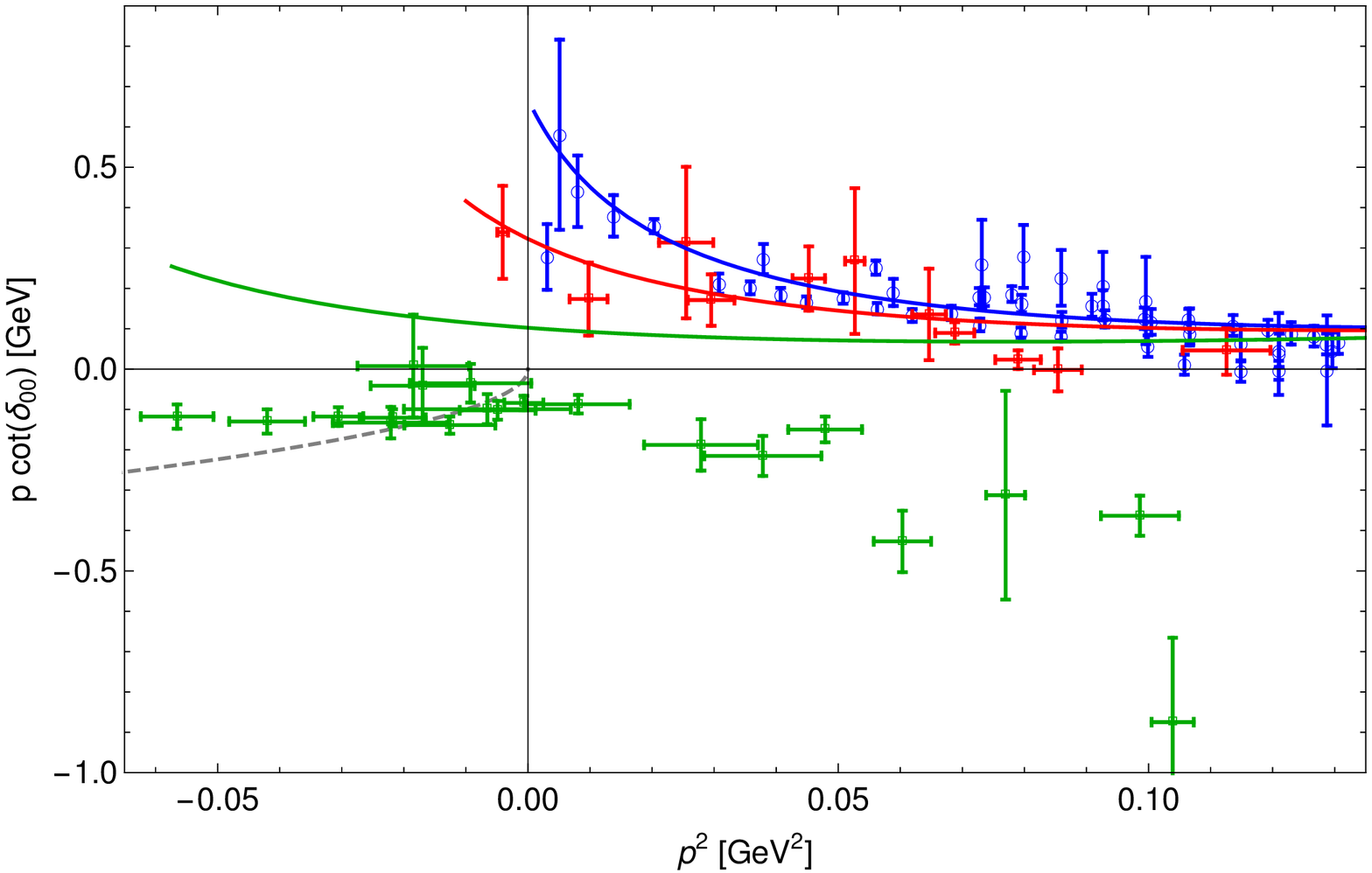}
\includegraphics[height=5.8cm]{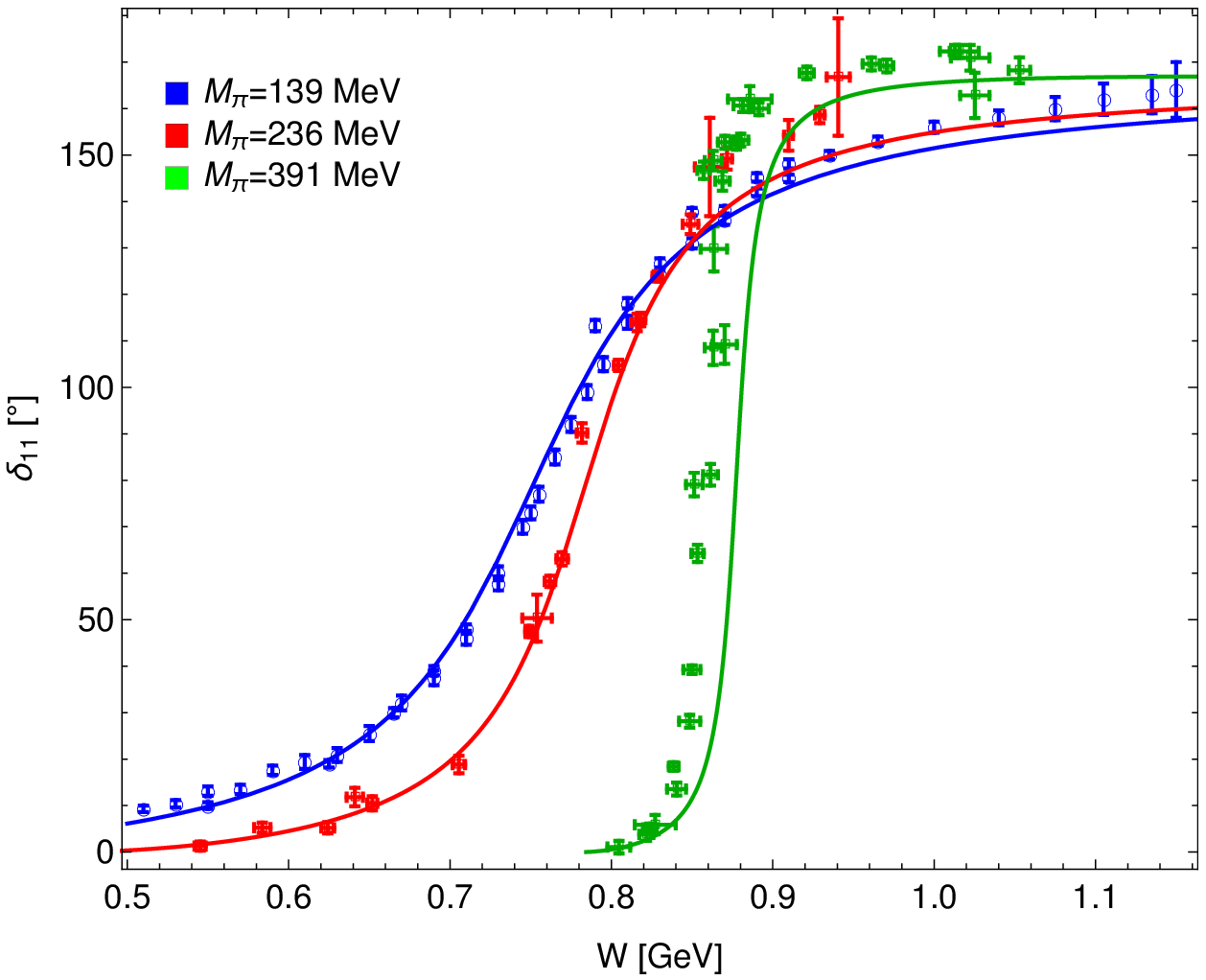}
\caption{Comparison of the chiral extrapolations using the mIAM with the LECs from Ref.~\cite{Hanhart:2008mx} to the lattice data from Ref.~\cite{Briceno:2016mjc,Wilson:2015dqa,Dudek:2012xn} (green and red data), whereas the experimental data from Refs.~\cite{Estabrooks:1974vu,Batley:2010zza,Hyams:1973zf,Protopopescu:1973sh,Grayer:1974cr,Gunter:1996ij} are shown in blue. The gray dashed line in the left figure shows the possible values of $p\,\cot(\delta_{00})$ for bound states, where $p$ denotes the modulus of the three-momentum in the center-of-mass $\pi\pi$ system.}\label{fig:PELprediction}
\end{figure}

Before analyzing the lattice data of the HadronSpectrum collaboration in the next section, it is instructive to see the predictions of different chiral unitary approaches at the pion masses in question. The data for $M_\pi=236$~MeV and $M_\pi=391$~MeV are shown in Fig.~\ref{fig:PELprediction} along with the experimental ones in red, green and blue, respectively. We predict them using (a) the 1-loop SU(2) mIAM amplitude with the values of LECs from Ref.~\cite{Hanhart:2008mx}; (b) The unitarized lowest-order chiral interaction; (c) the coupled-channel SU(3) IAM amplitude in the formulation and with the LECs of Refs.~\cite{Hu:2016shf, Guo:2016zos} that represents a slight modification of the original work of Ref.~\cite{Oller:1998hw}.

\textbf{(a)} In Ref.~\cite{Hanhart:2008mx} the free parameters $l_1^r$ and $l_2^r$ of the mIAM were fitted to reproduce all available experimental data at $\mu=770$~MeV, while keeping $l_3^r$ and $l_4^r$ fixed to the values of Ref.~\cite{Gasser:1984gg}. Using these LECs, see Tab.~\ref{tab:lecs}, we calculate the prediction of the mIAM for $\pi\pi$ scattering in the isoscalar and isovector channels for the physical as well as the two unphysical pion masses. The pion decay constant $F_\pi$ is calculated using the NLO chiral relation, Eq.~\eqref{eq:fpi} for the given $l_4^r$ and $F_\pi=92.4$~MeV at the physical pion mass. The result is depicted in Fig.~\ref{fig:PELprediction}, while the corresponding $\chi^2$ values per degree of freedom are collected in Tab.~\ref{tab:lecs}. Note that in the case of physical pion masses the latter does not reflect the full experimental data used in the original publication~\cite{Hanhart:2008mx}, but only the phase shifts in the quoted $\sigma$ and $\rho$ channels. The large value of the $\chi^2$ for the isovector channel (physical pion mass) is due to a yet unresolved conflict in the phase shifts extracted from experiment, see Refs.~\cite{Protopopescu:1973sh,Estabrooks:1974vu}. 

We observe in Fig.~\ref{fig:PELprediction} that the prediction of the mIAM works rather well for the light pion unphysical mass ($M_\pi=236$~MeV) both in the isoscalar and isovector channels. At higher pion mass (391 MeV), the extrapolation does rather well for the isovector case, but disagrees strongly with the recent lattice data in the isoscalar sector. Neither the sign of the scattering length, which is equal to the inverse of $p \cot(\delta_{00})$ at threshold, nor the presence of the bound state could be predicted. The position of the bound state can be read off Fig.~\ref{fig:PELprediction} as the intersection of the gray dashed line with the actual $p \cot(\delta_{00})$ prediction. As a matter of fact the presence of such a bound state was discussed in the original work on mIAM, Ref.~\cite{Hanhart:2008mx}, which, however, appeared there at much larger pion masses. This discrepancy might originate from two reasons: 1)~Based on the chiral expansion of the scattering amplitude to a finite order, the mIAM cannot have an infinite range of validity. This range might well be exhausted at pion masses as large as three times the physical one. If this is true, then the amplitude itself has to be modified, e.g., by including higher chiral orders~\cite{Pelaez:2010fj,Bijnens:1997vq}; 2)~In view of the rather good prediction in the isovector case, fixed through only two independent combinations of LECs, it may also be that including new (lattice) data shifts the LECs such that experimental and lattice data can be reconciled. For example, all four SU(2) LECs enter the determination of the isoscalar amplitude potentially leading to large correlations among them. Here, the additional information from the pion mass dependence will impose new and independent constraints, disentangling the correlations in the LECs. Here, we only consider this possibility and leave the inclusion of higher chiral orders in the interaction to future work.

\begin{figure}[t]
\includegraphics[height=5.7cm]{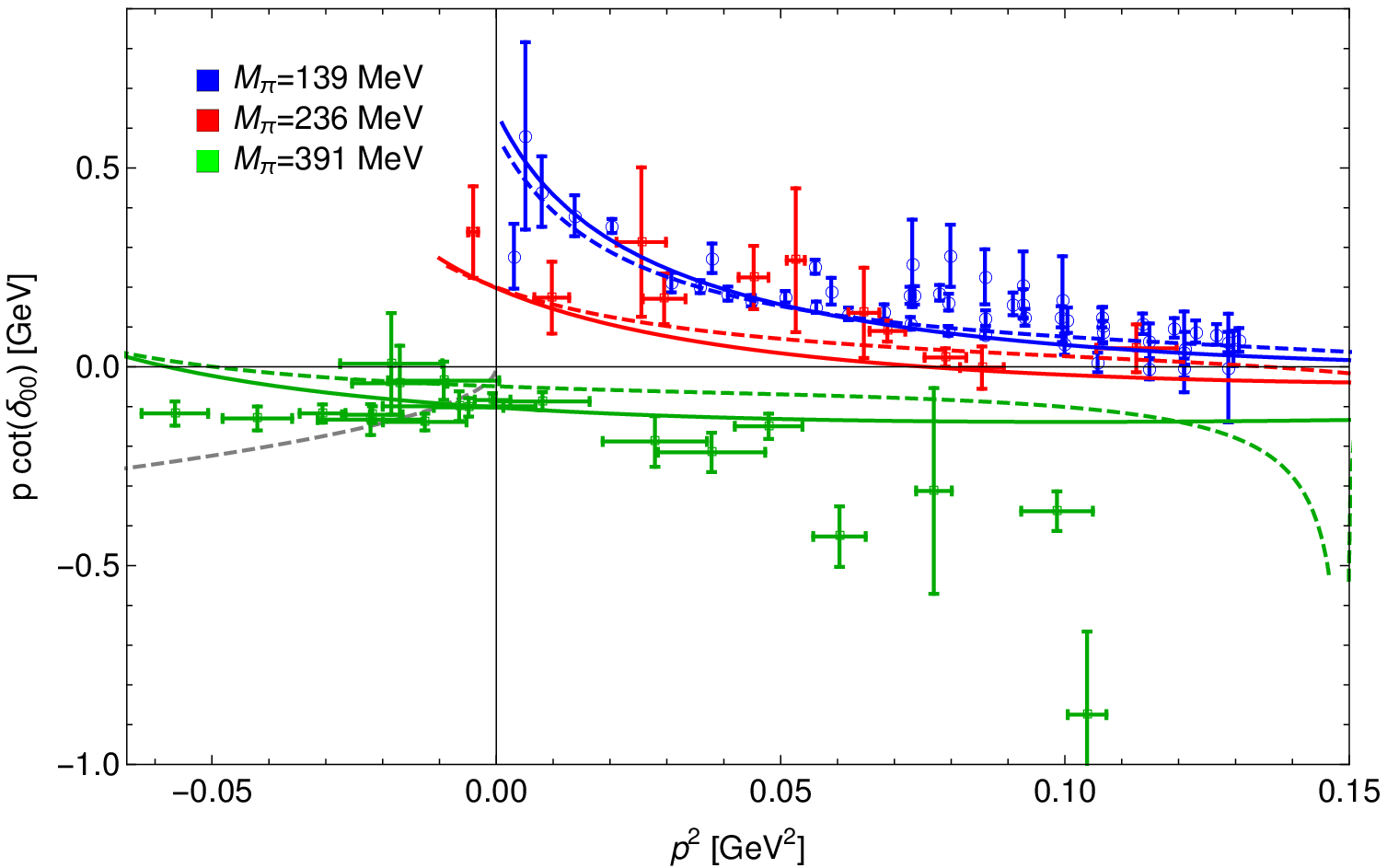}~
\includegraphics[height=5.7cm]{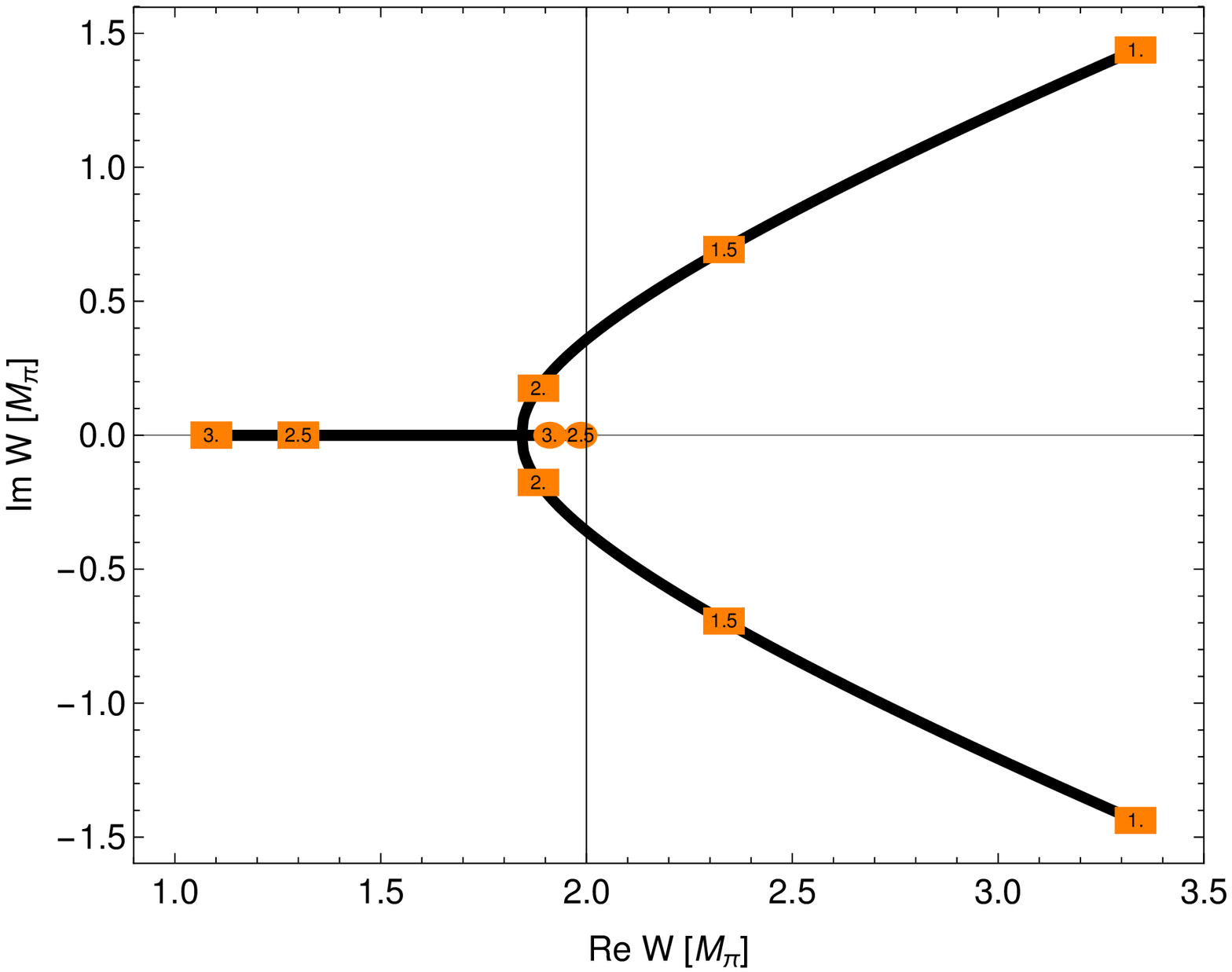}
\caption{[LEFT] Predictions of the LO BSE approach (solid lines) and the SU(3) model of Ref.~\cite{Hu:2016shf} (dashed lines). The same notation is adopted as in Fig.~\ref{fig:PELprediction}. [RIGHT] Behavior of the isoscalar resonance pole position with increasing pion mass for the LO BSE approach. The numbers denote the ratio of the pion mass to the physical one, whereas the shape of the numbering denotes the corresponding Riemann sheet - squares for the second and ellipses for the first one.}
\label{fig:SIGMA-OO}
\end{figure}

\textbf{(b)} As previously discussed, the isoscalar $\pi\pi$ amplitude from a re-summation of the lowest order (LO) chiral amplitude already leads to the generation of a pole for the $\sigma$~\cite{Oller:1997ti}. We therefore also consider the chiral prediction for this simplest of the considered amplitudes. We restrict the coupled-channel formalism of Ref.~\cite{Oller:1997ti} to a one-channel Bethe-Salpeter equation (BSE), which in the on-shell approximation reads 
\begin{align}
T_{\rm BSE}(E)=\frac{V(E)}{1-V(s)G(E,\Lambda)} \quad\text{ for }\quad 
V(E)=\frac{M_\pi^2-2E^2}{2 F_\pi^2}\,.
\end{align}
Here, $G(E)$ denotes the usual one-loop function evaluated with a cutoff, see Ref.~\cite{Oller:1997ti}, choosing it for the present qualitative discussion to be 1~GeV. Further, the pion decay constant is set equal to the physical one, which is allowed since the scattering amplitude is determined here only up to the leading chiral order. We obtain the chiral extrapolation shown in Fig.~\ref{fig:SIGMA-OO}. The corresponding predicted $\chi^2$ values are collected in Tab.~\ref{tab:lecs}. The LO BSE approach is capable to describe the lattice and experimental data fairly well, predicting also a bound state. With increasing pion mass the $\sigma$ resonance pole position in the complex energy plane changes. On the basis of mIAM as described in point \textbf{(a)} the corresponding trajectory was shown for the first time in Ref.~\cite{Hanhart:2008mx}.  The right panel of the Fig.~\ref{fig:SIGMA-OO} shows the  trajectory using the LO BSE approach for the pion mass increasing from 1 to 3 in units of the physical pion mass. As shown there, the poles move on the second Riemann sheet towards the real axis, becoming virtual states. After they meet at $M_\pi\sim 2.1$ in units of the physical pion mass, they split again to two poles, moving in opposite directions on the real axis. Finally, when one of the poles reaches the threshold at  $M_\pi\sim 2.5$ of the physical pion mass it disappears from the second and reappears on the first Riemann sheet as a bound state. Its binding energy grows then as the pion mass increases. Qualitatively, this pole trajectory shows the same features as described in Ref.~\cite{Hanhart:2008mx} for the mIAM approach, which, however, appear at much larger pion masses. For example, the jump to the first Riemann sheet occurs there at $M_\pi\sim 3.8$ of the physical pion mass, which is 140~MeV heavier than the heavy pion mass of the lattice simulation ($M_\pi=391$~MeV).
Note that the pole trajectory using the LO interaction is not very reliably determined due to the unknown dependence of the cut-off on the pion mass as well as the fact that the pion decay constant is actually changing for higher pion masses. Nevertheless, it is remarkable how similar the result of mIAM and LO BSE are at the physical pion mass whereas at higher pion masses they deviate strongly from each other. The lattice data actually favors the prediction of the LO approach. This suggests the necessity for a readjustment of the LECs used in the NLO mIAM.

\textbf{(c)} In Ref.~\cite{Hu:2016shf}, a model on the basis of the inverse amplitude method in three-flavor formulation with $\pi\pi$ and $K\bar K$ coupled channels was used to analyze various lattice QCD simulations of the $\rho$ channel. In the present work, we focus on the light quark sector only. However, the prediction of this model are instructive to understand at which energies the explicit dynamics from the kaon degrees of freedom becomes relevant.

The prediction of this model in the isoscalar channel is shown with the dashed curve in Fig.~\ref{fig:SIGMA-OO} to the left. We observe that at a pion mass of $236$~MeV or less the result is very similar to that of the LO BSE approach. In particular, a bound state for $M_\pi=391$~MeV is predicted. However, at the highest considered pion mass, a pronounced drop of $p\cot\delta$ at the larger values of $p^2$ appears. The reason is that in this scenario the $K\bar K$ threshold lies in the direct proximity of the considered energy region, namely at $p^2(2M_K)= 0.149\text{ GeV}^2$. We conclude that the lattice data at $M_\pi=391$~MeV and at high values of $p^2$ can only be analyzed when addressing the SU(3) effects properly, in particular the role of the $f_0(980)$ resonance. The four highest lattice data at $M_\pi=391$~MeV show a similar drop as the chiral prediction albeit at lower energies. We interpret this as the onset of the influence of the $K\bar K$ channel, possibly through the low-energy tail of the $f_0(980)$ that appears at lower energies in the lattice simulation than in the prediction.
Since the focus of the present work lies in the SU(2) sector only, in the fits to the data we will simply dismiss the last four data of the $M_\pi=391$~MeV lattice data. Indeed, we have found that it is impossible to find good SU(2) fits of the combined lattice data when excluding these points.

\section{Extrapolation from unphysical to physical pion masses}\label{sec:fit}

\begin{table}[t]
\begin{center}
\begin{tabular}{l ccc ccc cc}
\toprule\\
&&&&&&&\\
&&&&&&&\\
&&&&&&&\\
&&&&&&&\\
&&&&&&&\\
&&&&&&&\\
&&&&&&&\\
&
\begin{rotate}{90} mIAM from \cite{Hanhart:2008mx} \end{rotate}&
\begin{rotate}{90} LO BSE\end{rotate}&
\begin{rotate}{90} mIAM$^3_{\sigma(236,391)}$ \end{rotate}&
\begin{rotate}{90} mIAM$^3_{\sigma(236)\rho(236)}$ \end{rotate}&
\begin{rotate}{90} mIAM$^3_{\sigma(236,391)\rho(236,391)}$ \end{rotate}&
\begin{rotate}{90} mIAM$^4_{\sigma(236,391)}$ \end{rotate}&
\begin{rotate}{90} mIAM$^4_{\sigma(236)\rho(236)}$ \end{rotate}&
\begin{rotate}{90} mIAM$^4_{\sigma(236,391)\rho(236,391)}$ \end{rotate}\\
 \midrule
 $l_1\cdot10^3$	 	&$-3.7\pm0.2	$&--&$-14.2^{+1.7}_{-2.2}$&$-3.1^{~+0.2}_{~-0.2}  $&$~-2.6^{+0.0}_{-0.1} $&$-14.8^{+0.9}_{-2.5}$&$~-3.5^{~+0.3}_{~-0.2}  $&$~-2.6^{+0.1}_{-0.1}$\\
 $l_2\cdot10^3$ 	&$+5.0\pm0.4	$&--&$+21.7^{+3.2}_{-2.2}$&$+6.3^{~+0.5}_{~-0.5}  $&$~+7.9^{+0.0}_{-0.1} $&$+23.5^{+3.8}_{-1.0}$&$~+7.7^{~+0.6}_{~-0.8}  $&$~+8.6^{+0.0}_{-0.4}$\\
 $l_3\cdot10^3$ 	&$+0.8\pm3.8	$&--&$~-7.8^{+3.1}_{-2.9} $&$+5.3^{+10.5}_{-13.4} $&$-19.3^{~+0.5}_{~-0.0}$&$-12.6^{+4.4}_{-1.4}$&$~-8.9^{+17.3}_{-18.9}$&$-20.8^{+1.0}_{-0.0}$\\
 $l_4\cdot10^3$ 	&$+6.2\pm5.7	$&--&$+8.3~~		 $&$+8.3	          $&$~+8.3^{~~~}		      $&$-25.5^{+0.0}_{-0.4}$&$-29.9^{+10.2}_{~-7.3}$&$-17.7^{+3.0}_{-0.0}$\\
 \midrule
 $\chi^2_{\sigma(139)}$	&4.3	&4.8*	&4.0*	  &8.2*		&38.9*	&3.6*	  &5.4*	 	&30.1*	\\
 $\chi^2_{\rho(139)}$  	&9.4 	&--	&--	  &14.4*	&69.5*	&--	  &8.9* 	&7.5*	\\
 $\chi^2_{\sigma(236)}$	&2.9*	&1.9*	&1.6	  &0.8		&5.7	&1.4	  &0.9	 	&5.3	\\
 $\chi^2_{\rho(236)}$   &4.6*	&--	&--	  &4.2	  	&8.3	&--	  &1.9	  	&2.3	\\
 $\chi^2_{\sigma(391)}$	&46.8*	&1.6*	&1.6	  &39.0*	&3.8	&1.6	  &29.7*	&4.0	\\
 $\chi^2_{\rho(391)}$	&33.5*	&--   	&--	  &38.1*	&2.0	&--	  &95.5*	&1.4	\\
\bottomrule
\end{tabular}
\end{center}
\caption{Comparison of low-energy constants (upper part) and $\chi^2_{\rm d.o.f.}$ (lower part) in parameterizations described in the main body of the manuscript. The $\chi^2$ values are marked with a (*) when the corresponding data were predicted, i.e., not the subject of minimization.}
\label{tab:lecs}
\end{table}

\begin{figure}[t]
\includegraphics[width=0.55\linewidth]{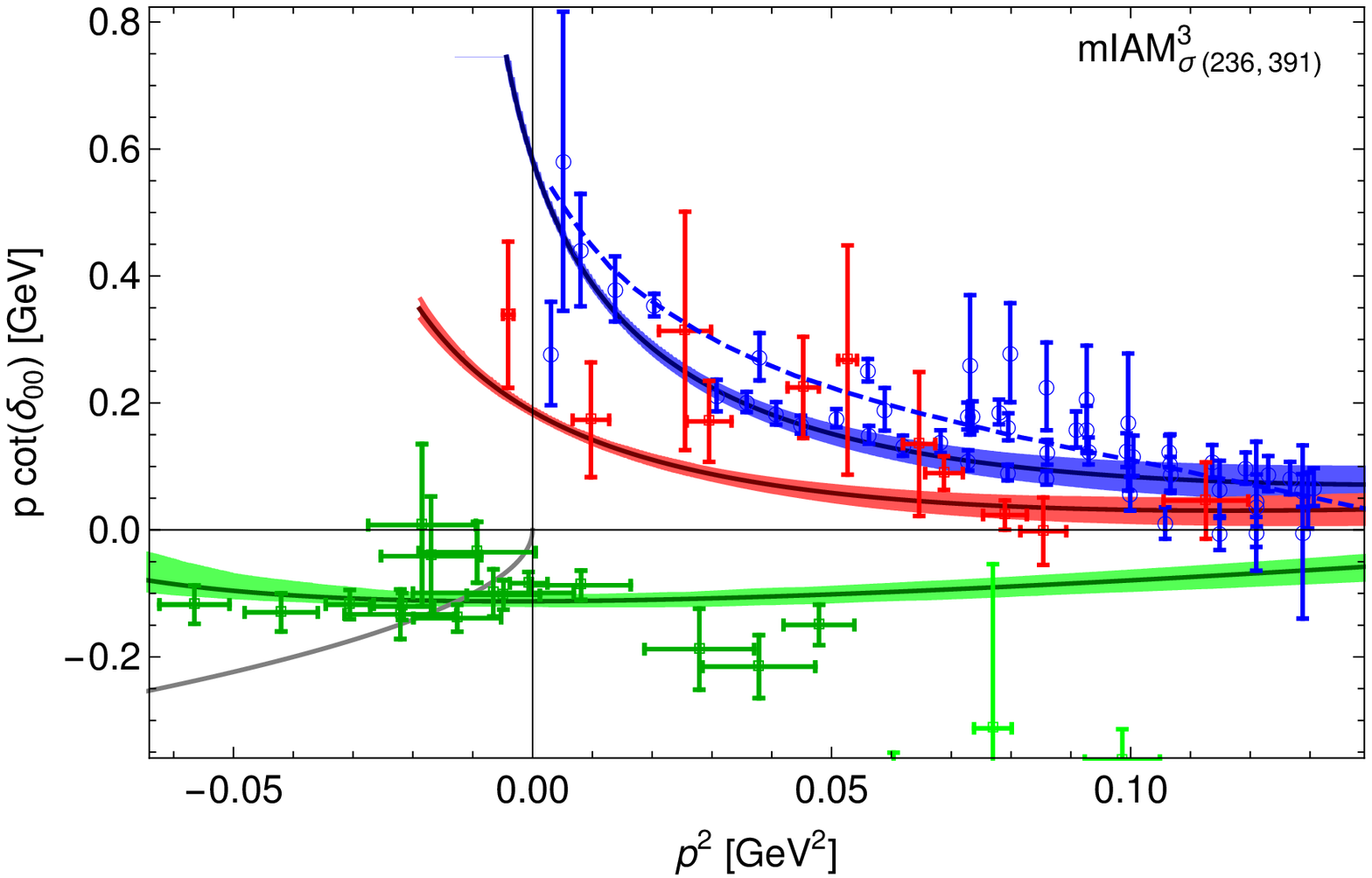}~
\includegraphics[width=0.43\linewidth]{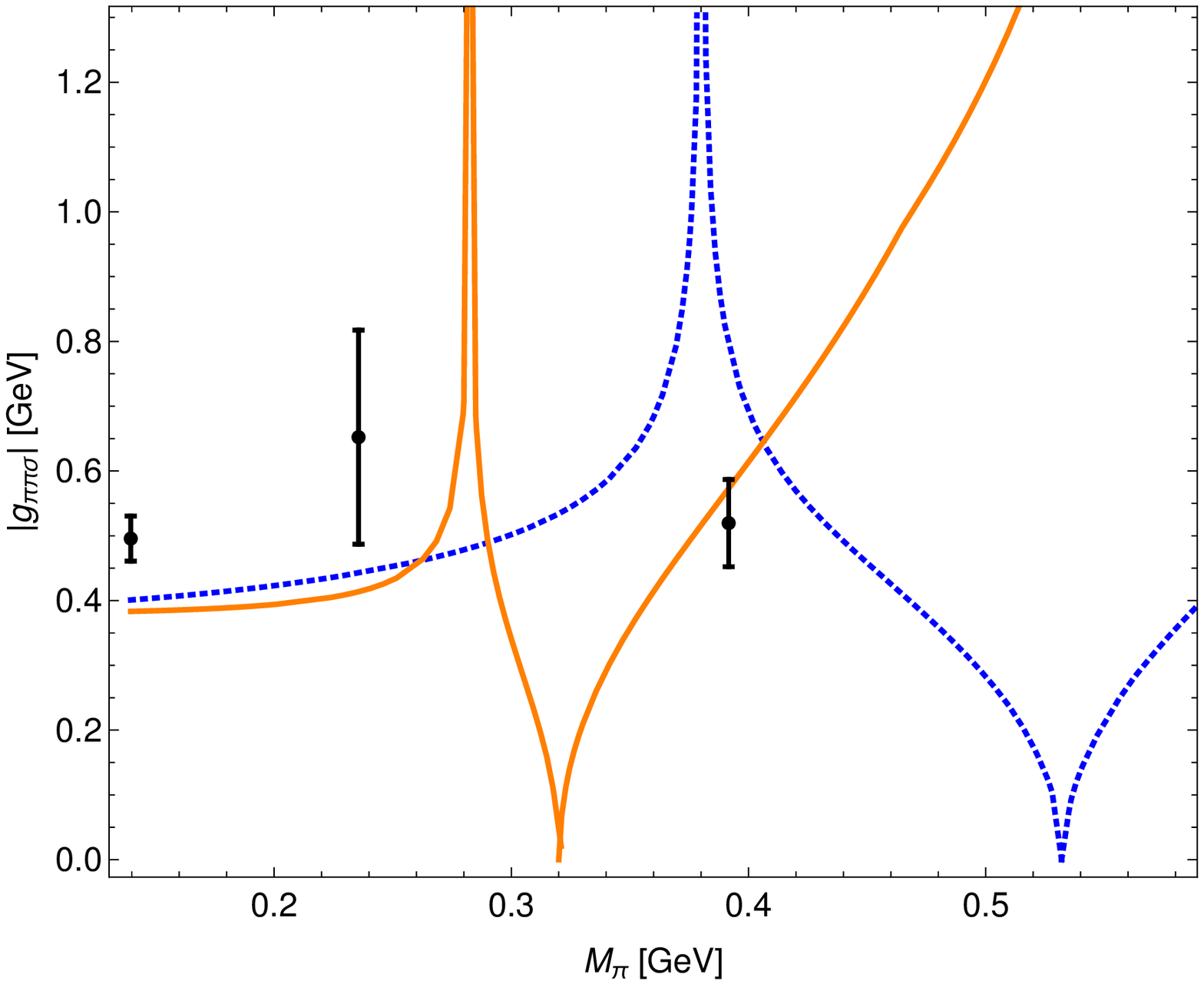}
\caption{[LEFT] Outcome of the mIAM approach when fitting the LECs to the lattice data~\cite{Briceno:2016mjc} at $M_\pi=236$ and $391$~MeV in the isoscalar channel only. The corresponding LECs and $\chi^2_{\rm d.o.f.}$ are collected in Tab.~\ref{tab:lecs} The experimental data (blue points) is the same as in Fig.~\ref{fig:PELprediction}, while the blue dashed line shows the outcome of the $S$-wave analysis of Ref.~\cite{Kaminski:2001hv}. [RIGHT] Strength of the $\sigma\pi\pi$ coupling (residuum of the pole) in the mIAM$^3_{\sigma(236,391)}$ scenario (solid orange curve) in comparison to the prediction of Ref.~\cite{Hanhart:2008mx} (dashed blue line) and various parametrization of Ref.~\cite{Briceno:2016mjc} as well as the dispersive determination of Ref.\cite{Pelaez:2010fj} at the physical pion mass (black dots with error bars).}\label{fig:bestsigma}
\end{figure}

\begin{figure}[p!]
\includegraphics[width=0.95\linewidth]{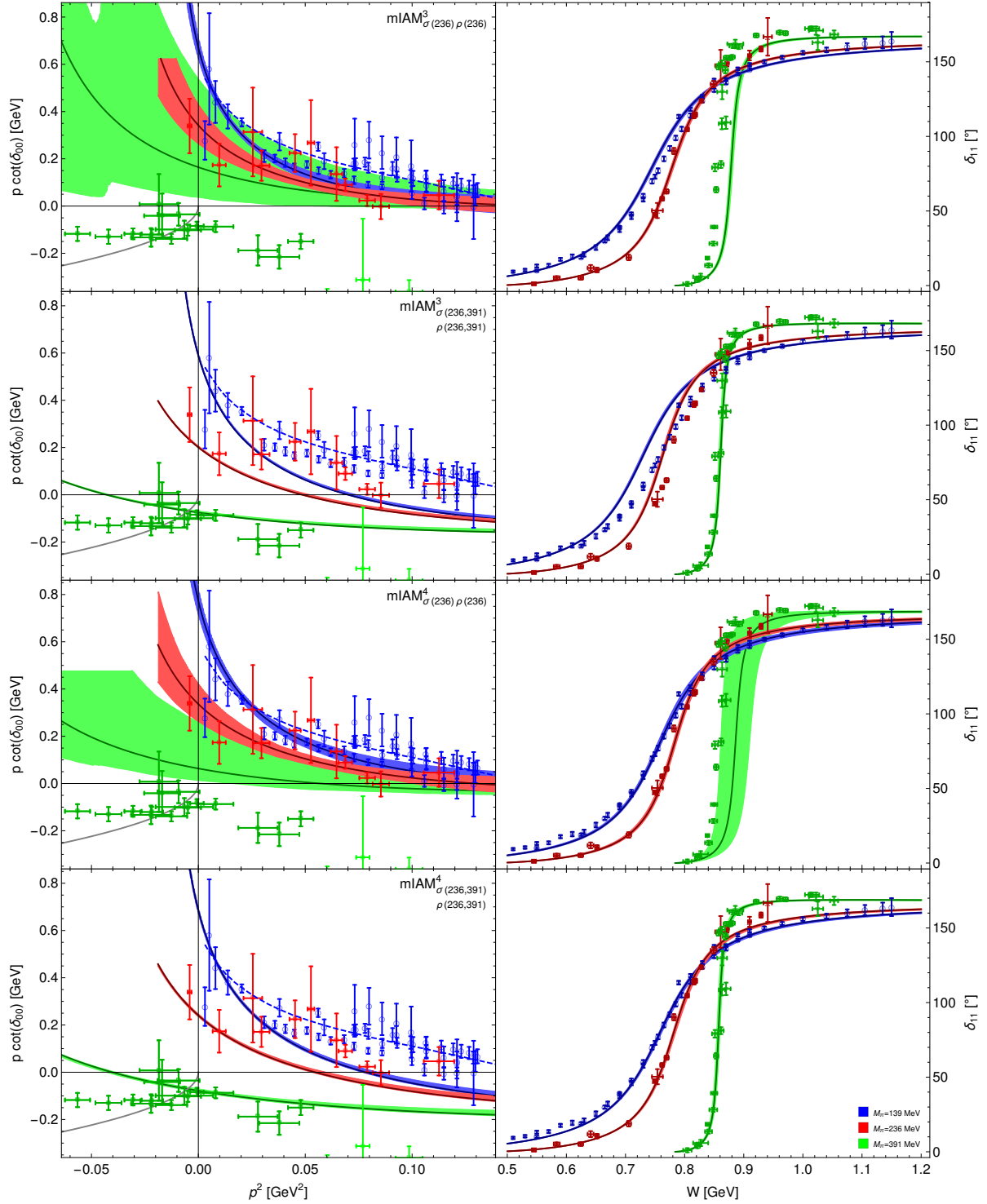}
\caption{Outcome of the mIAM approach in all considered fit scenarios, which are adjusted to reproduce the isovector(right column) and isoscalar(left column) data simultaneously. The corresponding LECs and $\chi^2_{\rm d.o.f.}$ are collected in Tab.~\ref{tab:lecs}. The red and green data points show the lattice data of Refs.~\cite{Wilson:2015dqa,Dudek:2012xn,Briceno:2016mjc} at $M_\pi=236$ and $391$~MeV, respectively. The experimental data from Refs.~\cite{Estabrooks:1974vu,Batley:2010zza,Hyams:1973zf,Protopopescu:1973sh,Grayer:1974cr,Gunter:1996ij} are represented by the blue points. The blue dashed line shows the outcome of the $S$-wave analysis of Ref.~\cite{Kaminski:2001hv}.
}\label{fig:ALLRHO}
\end{figure}

In the previous section we have shown that there is some tension between the lattice results on $\pi\pi$ scattering in the isoscalar channel~\cite{Briceno:2016mjc} and the chiral extrapolation based on the modified inverse amplitude method (mIAM) of Ref.~\cite{GomezNicola:2007qj} with LECs fixed to reproduce experimental data in Ref.~\cite{Hanhart:2008mx}. It is further shown that a re-summation of the leading order chiral amplitude already leads to quite a decent chiral extrapolation, predicting the presence of a bound state for a lattice setup with $M_\pi=391$~MeV. Therefore, the question is what new constraints on the values of the low-energy constants can be put when including new lattice data.

There are different scenarios, which we can test using the data of Refs.~\cite{Dudek:2012xn,Wilson:2015dqa,Briceno:2016mjc} addressing the following issues:
\begin{itemize}
 \item[1)] The $\pi\pi$ scattering phases are extracted from simulations performed at two different pion masses, i.e. $M_\pi=236$ and $391$~MeV. Using either one or both sets we can test the range of applicability of the mIAM. From the discussion in the previous section and the result shown in Fig.~\ref{fig:PELprediction} we expect that  at least the lighter pion mass setup is well within this range;
 \item[2)] At not too large pion masses the $\pi\pi$ interaction in the isovector channel is rather small at the leading chiral order. Thus, higher order terms are required to reproduce the $\rho$-resonance via re-summation techniques. We expect that whether both ($\sigma$ and $\rho$) or one channel ($\sigma$ only) is used as input for the fit, will have sizable effect on the obtained LECs, or, at least, their uncertainties;
 \item[3)] Pion decay constants at unphysical pion masses have been determined in Ref.~\cite{Hu:2016shf} using the fit of Ref.~\cite{Nebreda:2010wv} that well describes $F_\pi$ from lattice measurements up to at least $M_\pi\approx 400$~MeV. Using these decay constants as well as the NLO chiral formula Eq.~\eqref{eq:fpi} one of the LECs can be constrained to be $l^r_4=+0.0083$ reducing the number of parameters from 4 to 3. Note that this value is well within the error bars of the older determination~\cite{Gasser:1984gg}, which was used in the mIAM of Ref.~\cite{Hanhart:2008mx};
\end{itemize}
In the following, we will refer to the eight fit scenarios as mIAM$^{n}_{\sigma(\ldots)\rho(\ldots)}$ where $n$ indicates the number of free fit parameters (3 or 4) and the subscript denoting the channel ($\sigma, \,\rho$) and pion mass of the fitted data. In all cases, the fits are performed by minimizing the $\chi^2$ using the central values of Ref.~\cite{Hanhart:2008mx} as starting points. As argued before, the four data at the highest energies of the $391$~MeV data set are omitted. For the fit, the correlations between energies $W$ and phase shifts $\delta(W)$, or $p(W)\cot\delta(W)$, are given by the L\"uscher formalism and are taken into account. For the energy eigenvalues of the $\rho$ channel, the correlations among the energy eigenvalues themselves were available and were taken into account through correlated $\chi^2$ fitting.

The resulting LECs and the corresponding best $\chi^2$/d.o.f. values are collected in Tab.~\ref{tab:lecs}. The error bars in all our results were determined as follows: First, a large number of ensembles of fully re-sampled lattice data sets were generated taking into account the above-mentioned correlations. Then, starting from the best fit parameters, each re-sampled data set was refitted. Finally, uncertainties on LECs and phase shifts were determined from that set of refits. Both LECs and phase shifts (at a given fixed energy) sometimes exhibit very non-Gaussian distributions in their bootstrap samples. Instead of determining the variance from bootstrap samples it is then more meaningful to cut off the lower and higher ends of the distribution to obtain the 68\% confidence interval represented by the bands for the phase shifts and the errors for the LECs. Due to the high degree of non-linearity in the fit, sometimes the best fit lies almost at the boarder of the confidence interval as Tab.~\ref{tab:lecs} shows.

We have first tried to fit the data in the isoscalar channel at $M_\pi=236$~MeV only. However, the LECs acquired very unnatural values and the chiral extrapolation to the physical point was not satisfying. The reason lies in too much freedom for the fit function compared to the number of data points. This \textit{overfitting} happens with and without fixing $l^r_4$. Therefore, we desist from further discussions of this fit scenario. 

Furthermore, it is notable that when $l_4^r$ is used as a free parameter, its value tends to be negative in all fit scenarios as Tab.~\ref{tab:lecs} shows. Similarly, large negative values of $l^r_4$ have been found by fitting the isovector channel in Ref.~\cite{Bolton:2015psa}, i.e. $l_4^r=-28\cdot10^{-3}$. However, this is in conflict with the fact that the pion decay constant is a monotonically rising function of the pion mass, see, e.g., Refs.~\cite{Boucaud:2008xu,Beane:2007xs}. In view of Eq.~\eqref{eq:fpi} $l_4^r$ has to be positive in the relevant pion mass range. Thus, the 3 parameter fits can be considered as more consistent.

Fitting the $\sigma(236)$ and $\sigma(391)$ data leads to quite decent $\chi^2$ values as shown in the fourth and seventh column of Table~\ref{tab:lecs}, both for the fitted data and the chiral prediction at the physical point. However, the corresponding LECs are unnaturally large for both the three- and four-parameter scenarios. At this point it is worth mentioning that we tried different strategies to restrict the LECs in the fits to be close to the values of Ref.~\cite{Hanhart:2008mx}, resulting always in unsatisfactory descriptions of the lattice data in the $\sigma$ channel. This is a problem because the IAM coincides with the perturbative chiral expansion up to the next-to-leading order. Therefore, when the pion mass as well as the energy are not too high, the LECs are expected to be close to their standard ChPT values, see Ref.~\cite{Aoki:2013ldr}. On the other hand, it is clear that due to the bound state, observed close to the threshold in the $\sigma(391)$ setup, the effects renormalizing the usual chiral LECs can be enhanced. In conclusion, the fits mIAM$^{3/4}_{\sigma(236,391)}$ allow one to reconcile the lattice data at two different masses and to deliver a good chiral post-diction of experimental phase shifts as shown in Fig.~\ref{fig:bestsigma} to the left, but at the price of rather large LECs. 

In summary, the heavy lattice data in the $\sigma$ channel obviously lead to problems in the size of the LECs, and the fit to only the light data is not well constrained due to large uncertainties. The next logical step is therefore to include data from other quantum numbers in $\pi\pi$ scattering at light masses. 
The HadronSpectrum collaboration has also extracted phase shifts for the $I=L=1$ channel at the same pion masses of $M_\pi=236$ and $M_\pi=391$~MeV~\cite{Wilson:2015dqa,Dudek:2012xn}. As a first test, we have fitted the $\rho$-data at $M_\pi=236$~MeV only, resulting in very similar values and correlations of LECs as in Ref.~\cite{Bolton:2015psa}.

Next, we fit the lattice phase shifts at $M_\pi=236$ for both the $\sigma$ and the $\rho$ channels simultaneously with three parameters. This indeed stabilizes the LECs at natural values not too far away from the ones of Ref.~\cite{Hanhart:2008mx} as shown in Tab.~\ref{tab:lecs} in the fifth column. The first row in Fig.~\ref{fig:ALLRHO} shows the best fit (solid red lines for the data at $M_\pi=236$) and the predictions for the physical and the heavy pion masses (blue and green solid lines, respectively). While the prediction of the experimental phase shifts in the $\sigma$ channel is good, it is not satisfactory in the $\rho$ channel. One can understand this by noting that, with $l_4^r$ fixed, the only free parameter in the $\rho$-channel is given by the combination $-2l_1^r+l_2^r$. In all our fits to $\rho$ and $\sigma$ data, this combination is very similar to the value obtained in the fit of Ref.~\cite{Bolton:2015psa} to the isovector channel only, i.e. $(-2l_1^r+l_2^r)=14.7$.

Next, one can consider releasing $l_4^r$ in the combined fit of $\rho$ and $\sigma$ data at $M_\pi=236$~MeV, to remedy the above-mentioned problem that in the $\rho$ channel one has only one fit parameter. The outcome of this 4-parameter fit is indicated in the eighth column of Tab.~\ref{tab:lecs} and in the third row of Fig.~\ref{fig:ALLRHO}. The prediction of the experimental phase shifts improves drastically both in the $\rho$ and in the $\sigma$ channel. However, inspecting the values of the LECs it becomes clear that $l_4^r$ has acquired a large negative value of almost $-30\cdot 10^{-3}$ which is in fact quite close to the value of Ref.~\cite{Bolton:2015psa} as mentioned above. It seems that such a large negative value is always required to obtain a good fit and good chiral prediction in the $\rho$ sector (see also the  fit in the last row of Fig.~\ref{fig:ALLRHO}). 

Finally, one can include the lattice phase shift at the heavy pion mass ($M_\pi=391$~MeV) for fitting both the $\sigma$ and the $\rho$ channel to study the consistency of the model with data. The three-parameter fit exhibits a deteriorated $\chi^2$ of the fitted data and a rather bad post-diction of experimental data (second row of Fig.~\ref{fig:ALLRHO}). If $l_4^r$ is released the fit and prediction in the $\rho$ sector are quite good while the fit of the $\sigma$ phase shifts at $M_\pi=236$ is worse than in other considered cases, and so is the prediction of the experimental phase shifts. The small number of lattice phase shifts in combination with rather large uncertainties might be responsible for this, giving the  $\sigma$ phase shifts at $M_\pi=236$ simply not enough weight in the $\chi^2$.

	Comparing fit results for mIAM$^n_{\sigma(236)\rho(236)}$ with mIAM$^n_{\sigma(236,391)\rho(236,391)}$ for both $n$, the systematic limitation of the range of validity of the mIAM becomes even more evident. Seemingly a good fit to the $391$~MeV data can only be obtained sacrificing the agreement with the experimental data. Or stated differently, if the prediction at the physical point agrees with the experimental data, it differs significantly from the lattice data at the highest pion mass.

\begin{table}[t]
\begin{center}
\begin{tabular}{|l| lll| lll| lcl| }
\hline
	& \multicolumn{3}{c|}{$M_\pi=139$~MeV}
	& \multicolumn{3}{c|}{$M_\pi=236$~MeV}
	& \multicolumn{3}{c|}{$M_\pi=391$~MeV}\\
\cline{2-10}
Approach	
	& Re $W$&-Im $W$&$|g_{\sigma\pi\pi}|$
	& Re $W$&-Im $W$&$|g_{\sigma\pi\pi}|$
	& Re $W$&-Im $W$&$|g_{\sigma\pi\pi}|$\\
\hline
mIAM~\cite{Hanhart:2008mx}
		&449	&218	&401
		&524	&175	&451	
		&--	&--	&--\\
LO BSE
		&$465			$&$201		$&$416			$
		&$501			$&$117		$&$428			$
		&$761			$&$0		$&$486			$\\
Disp. Rel. ~\cite{Pelaez:2015qba}
		&$449_{-16}^{+22}	$&$275_{-12}^{+12}	$&$495^{+35}_{-35}	$
		&--	&--	&--
		&--	&--	&--\\
K-Matrix~\cite{Briceno:2016mjc}
		&--&--&--
		&$667_{-113}^{+113}	$&$201_{-84}^{+84}	$&$652^{+165}_{-165}	$
		&$753_{-8}^{+8}		$&$0			$&$520_{-68}^{+68}	$\\
\hline
mIAM$^3_{\sigma(236,391)}$ 
		&$434^{+4}_{-6}		$&$197^{+8}_{-7}	$&$383^{+1}_{-3}	$
		&$484^{+5}_{-6}		$&$117^{+2}_{-9}	$&$414^{+4}_{-10}	$
		&$750^{+5}_{-4}		$&$0			$&$573^{+17}_{-37}	$\\
mIAM$^3_{\sigma(236)\rho(236)}$ 
		&$459^{+4}_{-6}		$&$181^{+14}_{-12}	$&$391^{+4}_{-6}	$
		&$549^{+23}_{-68}	$&$148^{+21}_{-3}	$&$441^{+2}_{-10}	$
		&-- 			 &--			 &--			\\
mIAM$^3_{\sigma(236,391)\rho(236,391)}$ 
		&$452^{+1}_{-0}		$&$144^{+0}_{-4}	$&$366^{+3}_{-0}	$
		&$515^{+0}_{-0}		$&$104^{+3}_{-0}	$&$421^{+2}_{-1}	$
		&$771^{+1}_{-0} 	$&$0			$&$423^{+1}_{-9}	$\\
\hline
\end{tabular}
\end{center}
\caption{Comparison of the $\sigma$ resonance properties (all in MeV) in various approaches and at different pion masses. The isoscalar pole of the $\pi\pi$ scattering is on the second Riemann sheet at $M_\pi=139$ and $236$~MeV, but on the first one at $M_\pi=391$~MeV (bound state). In the latter case "--" indicates that no bound state was found.}\label{tab:poles}
\end{table}

Finally, for unphysical pion masses we predict the pole position and its residue $|g_{\sigma\pi\pi}|$ that is a measure of the $\sigma$ coupling strength to $\pi\pi$. Specifically, we address the discrepancy of the mIAM chiral predictions of $|g_{\sigma\pi\pi}|$ made in Refs.~\cite{Nebreda:2010wv,Nebreda:2011di,Hanhart:2008mx} and the analysis of Ref.~\cite{Briceno:2016mjc}. In Tab.~\ref{tab:poles} the results derived in Ref.~\cite{Briceno:2016mjc} from fits to the lattice data on the $\sigma$ using different parameterizations, are compared with the outcome of the approaches discussed in this work and with the most recent dispersive determination of the resonance parameters at physical pion masses~\cite{Pelaez:2015qba}. The error bars on our predictions were determined again via re-sampling. The $M_\pi$ dependence of the pole residuum $|g_{\sigma\pi\pi}|$ is depicted in the right part of Fig.~\ref{fig:bestsigma} (solid orange line), considering the extrapolation mIAM$^{3}_{\sigma(236,391)}$, together with the prediction using the LECs of Ref.~\cite{Hanhart:2008mx} (blue dashed line) as well as the outcome from the parameterizations discussed in Ref.~\cite{Briceno:2016mjc} and dispersive determination of Ref.~\cite{Pelaez:2015qba} (points with error bars). 

The movement of the isoscalar pole in the framework of the mIAM was discussed in Sec.~\ref{sec:pred} and is reflected in Fig.~\ref{fig:bestsigma} via three steps: 1) The residuum grows rapidly with the pion mass when the pole moves towards the real energy axis, see also right part of Fig.~\ref{fig:SIGMA-OO}. A first kink in $|g_{\sigma\pi\pi}|$ is observed when the pole reaches the real axis at $M_\pi\approx280$~MeV; 2) Now, the pole separates into two and one of the poles moves along the real energy axis upwards to the two-pion threshold (the residue of the other pole moving to smaller energies is not shown), with a decreasing residue that becomes zero at $M_\pi\approx320$~MeV; 3) At this pion mass the pole ``jumps'' from the second to the first Riemann sheet leading to the second kink in $|g_{\sigma\pi\pi}|$. With further increasing pion mass the strength of the pole increases again as the state becomes more bound. This behavior is common to mIAM as well as to the LO BSE approach. However, the scaling of the trajectory with the pion mass and the pion mass beyond which a bound state appears can only be determined by taking the recent lattice data of Ref.~\cite{Briceno:2016mjc} into account.
The pole and residue trajectories predicted here suggest that it would be very enlightening to measure the isoscalar phase shift in a range of pion masses between 280 and 320 MeV, to confirm or reject the unusual behavior of the $\sigma$ pole.

\section{Summary}\label{sec:sum}

We have tested the modified inverse amplitude method (mIAM), which matches with the chiral expansion up to the next-to-leading order, against lattice QCD data in the isoscalar and isovector channels of two-pion scattering that were recently determined by the HadronSpectrum collaboration. 
Our analysis is preliminary insofar that the lattice data in the isoscalar channel have not been peer-reviewed at the date of submission of this manuscript.
In the isoscalar channel a reconciliation of the lattice data at two different pion masses ($M_\pi=236$~MeV and 391~MeV) with the experimental phase shifts is possible within the mIAM. However, the resulting low-energy constants are unnaturally large. The inclusion of the isovector data in a simultaneous fit to phase shifts at $M_\pi=236$~MeV leads to smaller low-energy constants in the vicinity of the standard values from ChPT. However, in such fit scenarios the experimental data and the lattice data at the highest pion mass $M_\pi=391$~MeV cannot be reproduced simultaneously using the modified inverse amplitude method. 

Furthermore, we have analyzed the behavior of the $\sigma$ pole position and its strength as a function of the pion mass. Qualitatively, a similar behavior as discussed in Ref.~\cite{Briceno:2016mjc} was found. However, the imaginary part of the pole position as well as its strength are somewhat larger than in the ChPT inspired models, discussed in the present work. Our analysis suggests that the pole corresponding to the $\sigma$ resonance changes the Riemann sheet at roughly $M_\pi=320$~MeV. In this region the $\sigma\pi\pi$ coupling changes rapidly with the pion mass. Therefore, it would be specifically interesting to have lattice QCD data at this pion mass allowing for a critical test of ChPT inspired re-summation models.

~\\
\noindent
\textbf{Acknowledgements}
We thank P.~C.~Bruns and R. Brice\~no for useful discussions. The work of M.M. has been supported by the German Research Foundation (DFG) through the research fellowship No. MA~7156/1. 
M.D. is supported by the National Science Foundation (CAREER grant 
PHY-1452055 and PIF grant No. 1415459) and by the U.S. Department of Energy, Office of Science, 
Office of Nuclear Physics under contract DE-AC05-06OR23177 and grant No. DE-SC0016582.


\end{document}